\documentclass[11pt]{article}
\usepackage{moriond,epsfig}

\usepackage{multicol}
\def\Journal#1#2#3#4{{#1} {\bf #2}, #3 (#4)}

\def\be{\begin{equation}}
\def\ee{\end{equation}}
\def\bea{\begin{eqnarray}}
\def\eea{\end{eqnarray}}

\begin{document}
\vspace*{4cm}
\title{Radioscience simulations in General Relativity and in alternative theories of gravity}

\author{A. Hees$^{1,2}$, P. Wolf$^2$, B. Lamine$^3$, S. Reynaud$^3$, M.T. Jaekel$^4$, C. Le Poncin-Lafitte$^2$, V. Lainey$^5$, A. F\"uzfa$^6$, V. Dehant$^1$}

\address{$^1$ Royal Observatory of Belgium, Avenue Circulaire 3, 1180 Bruxelles, Belgium, ahees@oma.be \\
$^2$ LNE-SYRTE, Observatoire de Paris, CNRS, UPMC, France \\
$^3$ Laboratoire Kastler Brossel, UPMC, Campus Jussieu, France \\
$^4$ Laboratoire de Physique Th\'eorique de l'ENS, Paris, France \\
$^5$ IMCCE, Observatoire de Paris, France \\
$^6$ naXys, University of Namur (FUNDP), Belgium}

\maketitle\abstracts{In this communication, we focus on the possibility to test GR with radioscience experiments. We present a new software that in a first step simulates the Range/Doppler signals directly from the space time metric (thus in GR and in alternative theories of gravity). In a second step, a least-squares fit of the involved parameters is performed in GR. This software allows one to get the order of magnitude and the signature of the modifications induced by an alternative theory of gravity on radioscience signals. As examples, we present some simulations for the Cassini mission in Post-Einsteinian gravity and with the MOND External Field Effect.}

\section{Introduction}
There is still a great interest in testing General Relativity (theoretical motivations such as quantification of gravity, unification with other interactions\dots). Within the solar system, the gravitational observations are always related with radioscience measurements (Range and Doppler) or with angular measurements (position of body in the sky, VLBI). In this communication, we  present a new tool that performs simulations of radioscience experiments in General Relativity and in alternative metric theories of gravitation. The developed software simulates Range and Doppler signals directly from the space time metric (and from the initial conditions of the bodies considered). The aim of this tool is to provide orders of magnitude and signatures for the variations of the Range/Doppler induced by alternative theories of gravity. In order to compare signals in different theories, a least-squares fit of the different parameters involved in the problem (mainly the initial conditions) is performed. In this communication, we present the principles of the software and results of simulations of the Cassini probe mission during its cruise from Jupiter to Saturn with Post-Einsteinian modifications of gravity (PEG)~\cite{peg1,peg2} or with an external field effect due to a MONDian modification of gravity~\cite{mond}.

\section{Description of the software}
Our software directly generates Doppler/Range signal from the metric considered in a fully-relativistic way. This includes the integration of the equations of motion in a given coordinate time, the computation of the time transfer of light and the clock behavior.

The equations of motion are derived from the metric and the geodesic equations integrated with respect to coordinate time~\cite{misner}.
 
The connection between the coordinate and proper time of clocks is obtained by integrating the equation of proper time~\cite{misner} 
\begin{equation}
 \frac{d\tau}{dt}=\sqrt{g_{00}-2g_{0i}v_i-g_{ij}v_iv_j}
\end{equation}
where $g_{\mu\nu}$ is the space-time metric, $v_i$ is the velocity of the clock and $\tau$ its proper time.

Finally, the time transfer is also determined directly from the metric using Synge World's function formalism~\cite{synge}. Within this formalism, one does not need to integrate the photon trajectory in order to get the time transfer or the frequency shift (in the linear approximation). Instead, those quantities are expressed as integrals of functions defined from the metric (and their derivatives) along the photon Minkowski path. For example the coordinate propagation time can be expressed as~ \cite{synge}
\be \label{time}
T(x_e^i(t_e),x_r^i(t_r),t_r)=\frac{R_{er}}{c}+\frac{R_{er}}{c}\int_0^1f(z^\alpha(\mu))d\mu 
\ee
with
\be
f=-h_{00}-2N^i_{er}h_{0i}-N^i_{er}N^j_{er}h_{ij}, \qquad h_{\mu\nu}=g_{\mu\nu}-\eta_{\mu\nu}, \qquad N^i_{er}=\frac{x^i_r-x^i_e}{R_{er}},
\ee
$x^i_e,t_e$ are the position and time of the emitter (computed iteratively assuming flat space time), $x^i_r, t_r$ are the position and time of the receptor and $R_{er}=\left\|x^i_e(t_e)-x^i_r(t_r)\right\|$. The integral in (\ref{time}) is performed over a Minkowski path between emitter and receptor ($z^\alpha(\mu)$ is a straight line). A similar expression is used for the frequency shift.
 
In order to investigate the observable signatures of an alternative theory of gravity in the Range and Doppler data we perform a least-squares fit in GR on the different parameters (initial conditions and masses of the bodies) and search for identifiable signatures in the residuals. The fit of initial conditions is necessary in order to avoid effects due to the choice of coordinates. As a matter of fact, this fit is always done in practice, which means that a reasonable analysis of the signal produced by an alternative theory has to be done after the fit.

\section{Simulations of Cassini mission}
As an example, we simulate the two-way range and Doppler signals to the Cassini spacecraft from June 2002 during 3 years (when the probe was between Jupiter and Saturn). To simplify the situation we only consider the Sun, the Earth and Cassini spacecraft.

\subsection{Post-Einsteinian Gravity (PEG)}
The first alternative metric theory considered is Post-Einsteinian Gravity (PEG)~\cite{peg1,peg2}. From a phenomenological point of view this theory consists in including two potentials $\Phi_N(r)$ and $\Phi_P(r)$ to the metric. Here we concentrate on the sector $\Phi_P(r)$. We consider a series expansion. That is to say we suppose the spatial part of the metric to be modified as follows
\be
g_{ij}=\left[g_{ij}\right]_{GR} -2\delta_{ij} \left(\chi_1 r+\chi_2r^2+\delta\gamma\frac{GM}{c^2r}\right)
\ee
where $r$ is a radial isotropic coordinate, $M$ is the sun mass, $c$ the velocity of light and $G$ the gravitational constant. The parameter $\delta\gamma$ is related to the post-newtonian parameter $\delta \gamma=\gamma-1$.

\begin{figure}[h]
\psfig{figure=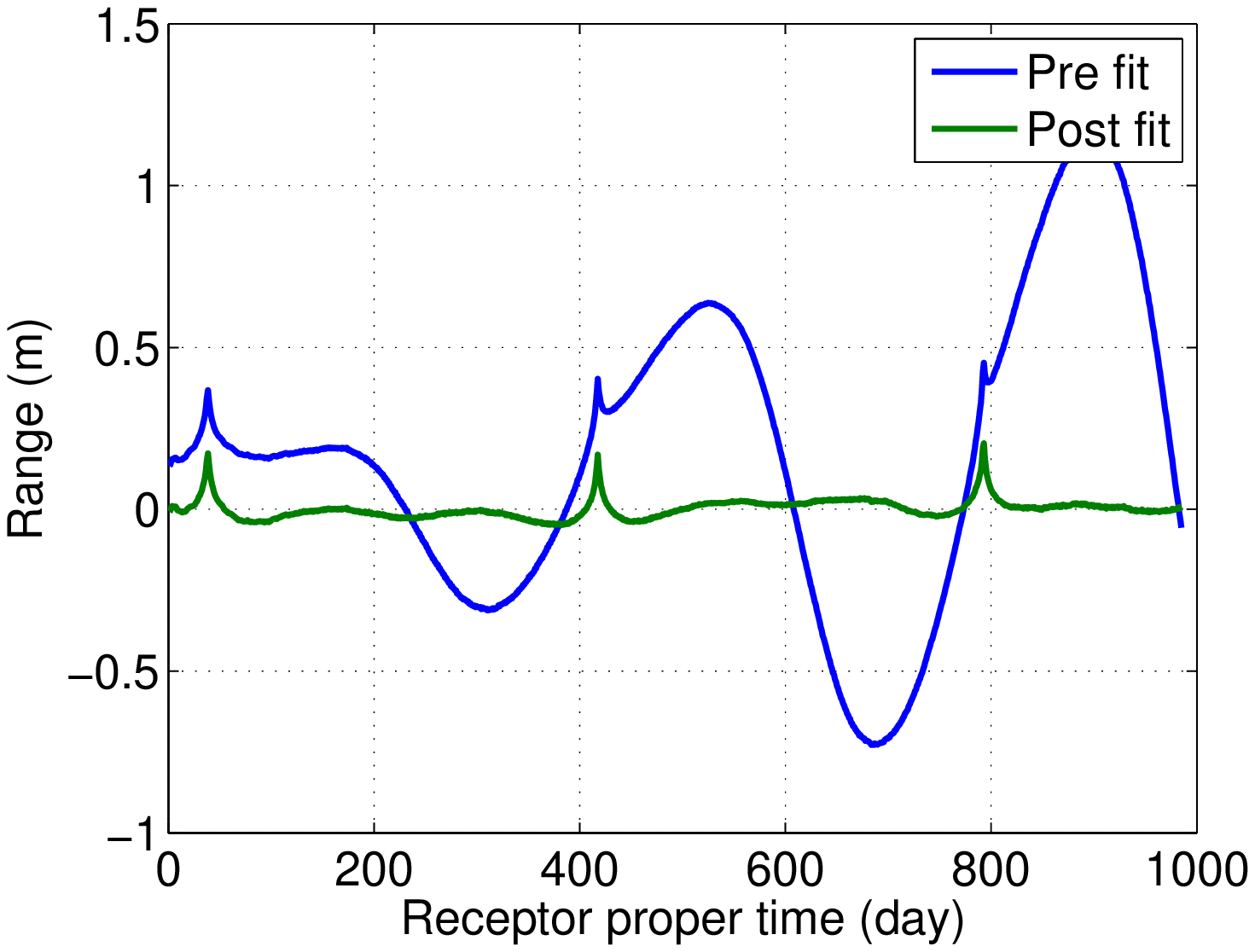,width=0.47\textwidth}\hfill
\psfig{figure=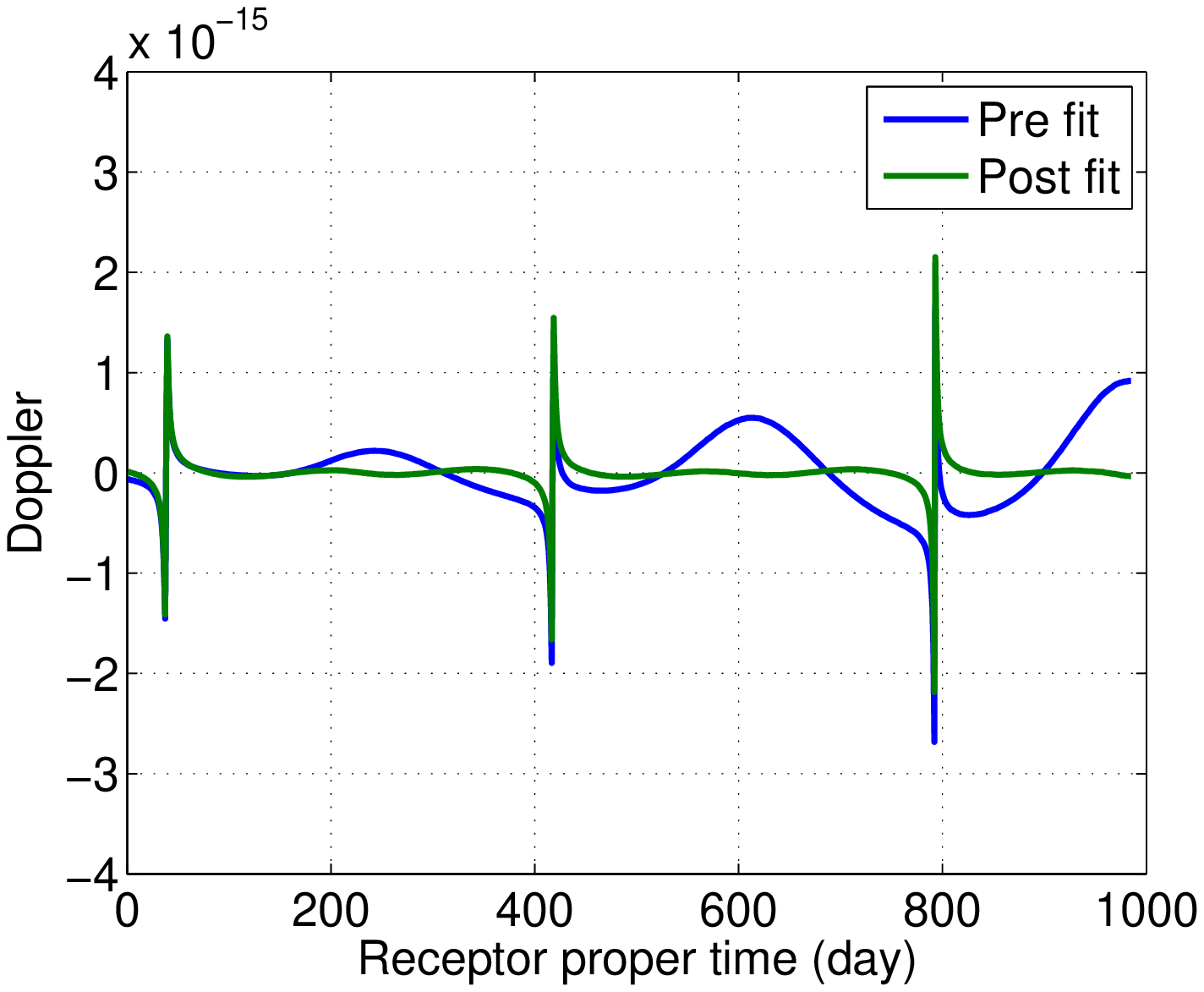,width=0.47\textwidth}
\caption{Representation of the Range (on the left) and Doppler (on the right) signals due to an alternative theory with $\gamma-1=10^{-5}$. The blue line is the difference between a simulation in the alternative theory and a simulation in GR (with the same parameters). The green line is the residuals obtained after analyzing the simulated data in GR (which means after the fit of the different parameters).}
\label{figDiff}
\end{figure}

Different simulations were performed with different values of the three PEG parameters. For example, Figure \ref{figDiff} represents the Range and Doppler differences between a simulation in a theory with $\delta\gamma=\gamma-1=10^{-5}$ and in GR. The three peaks occur during solar conjunctions. The signal due to the conjunction is not absorbed at all by the fit of the initial conditions which nevertheless absorbs a large modulations in the range signal.

\begin{figure}[h]
 \psfig{figure=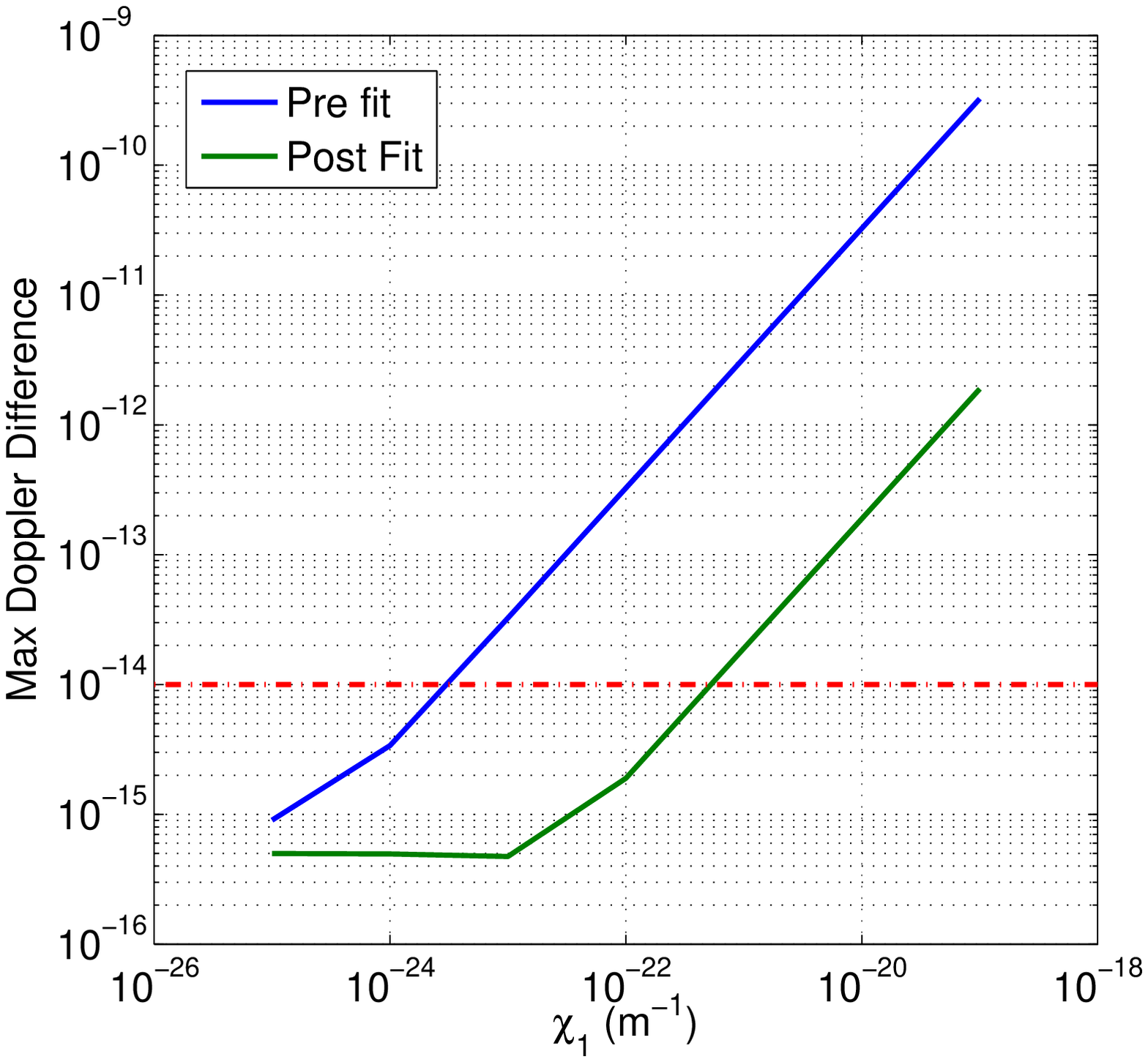,width=0.33\textwidth}
 \psfig{figure=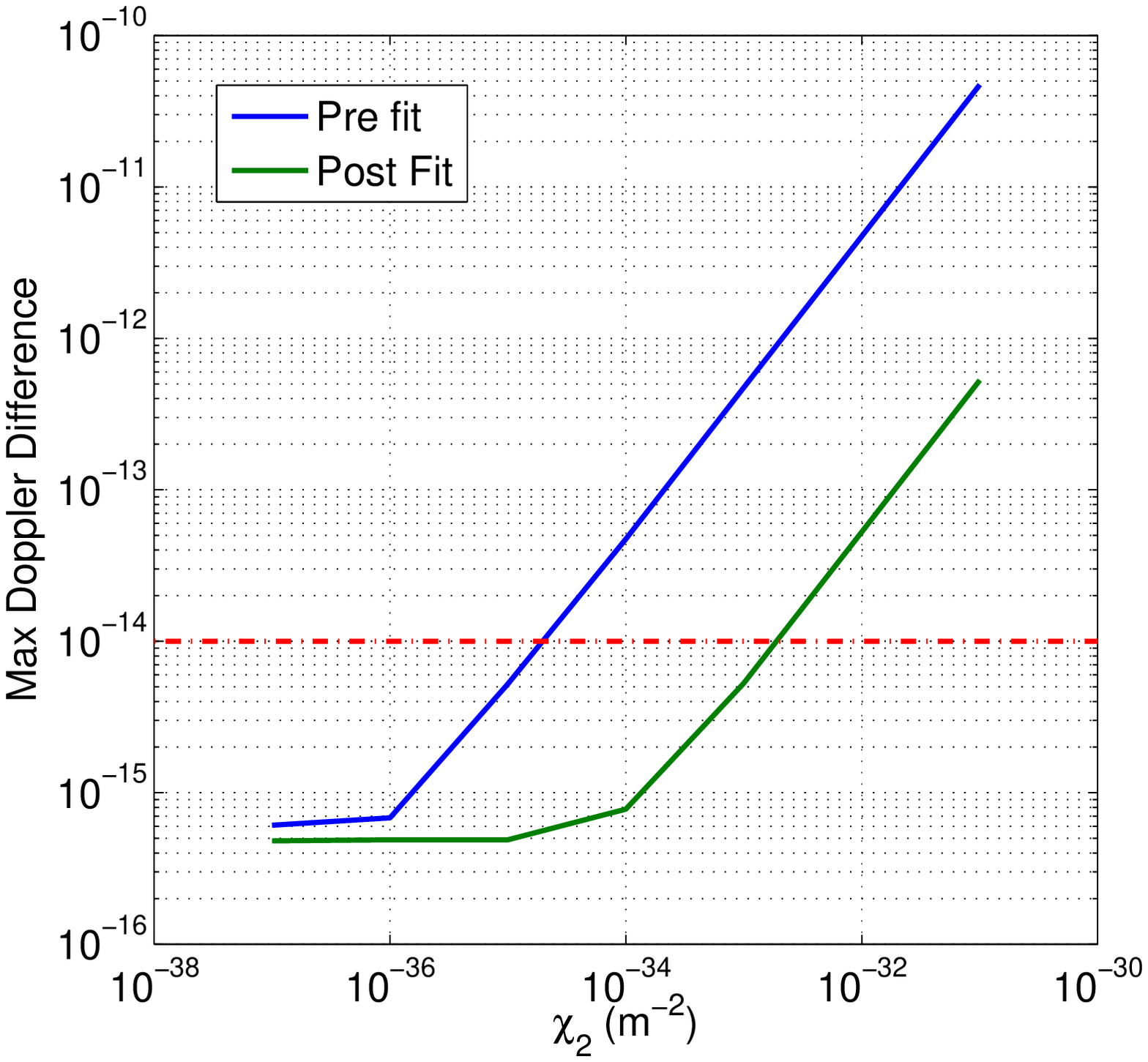,width=0.33\textwidth}
 \psfig{figure=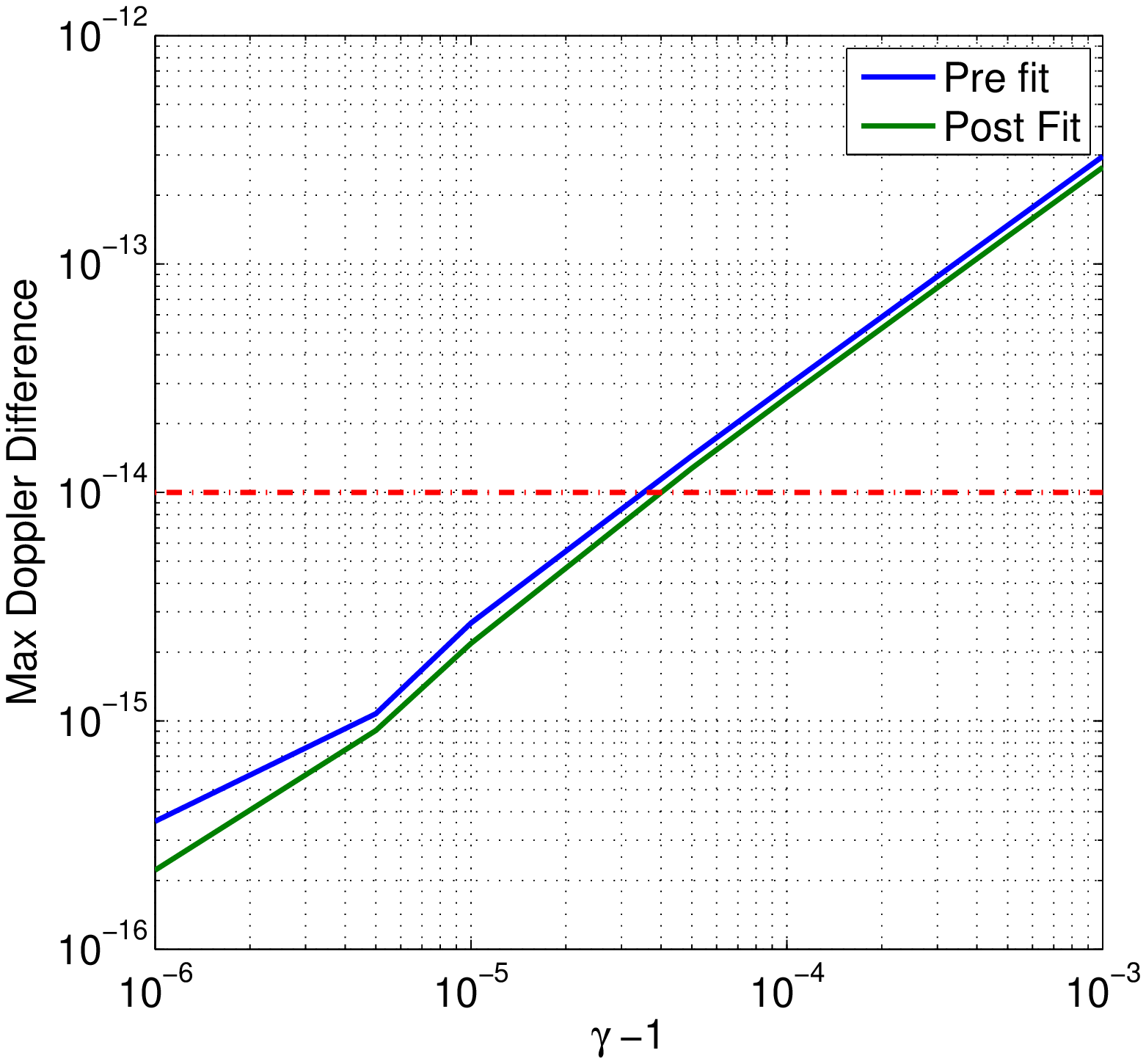,width=0.33\textwidth}
 \caption{Representation of the maximal Doppler signal due to PEG theory (parametrized by three parameters $\chi_1, \chi_2, \gamma-1$)  for the Cassini mission between Jupiter and Saturn. The blue lines represent the maximal Doppler difference between a simulation in the alternative theory and a simulation in GR (with the same parameters). The green lines represent the maximal residuals obtained after analyzing the simulated data's in GR (i.e. after the fit of the parameters). The read lines represent the assumed Cassini accuracy.}
 \label{figPEG}
\end{figure}

To summarize, Figure \ref{figPEG} represents the maximal difference between the Doppler generated in PEG theory and the Doppler generated in GR for different PEG theories (characterized by their values of $\chi_1$, $\chi_2$ and $\delta \gamma$). If we request the residuals to be smaller than Cassini Doppler accuracy (roughly $10^{-14}$), we get boundary values for the three parameters: $\chi_1<5\ 10^{-22}m^{-1}$, $\chi_2<2\ 10^{-33}m^{-2}$ and $\gamma-1<3 \ 10^{-5}$ (which is very similar to the real estimation~\cite{cassini}).

\subsection{MOND External Field Effect (EFE)}
Another alternative theory considered is the External Field Effect produced by a MOND theory~\cite{mond}. In this framework, the dominant effect is modeled by a quadrupolar contribution to the Newtonian potential $U=\frac{GM}{r}+\frac{Q_2}{2}x^ix^j\left(e_ie_j-\frac{1}{3}\delta_{ij}\right)$ where $e_i$ is a unitary vector pointing towards the galactic center and $2.1 \ 10^{-27}s^{-2}\leq Q_2\leq 4.1 \ 10^{-26}s^{-2}$ is the value of the quadrupole moment whose value depends on the MOND function. 

\begin{figure}[h]
\psfig{figure=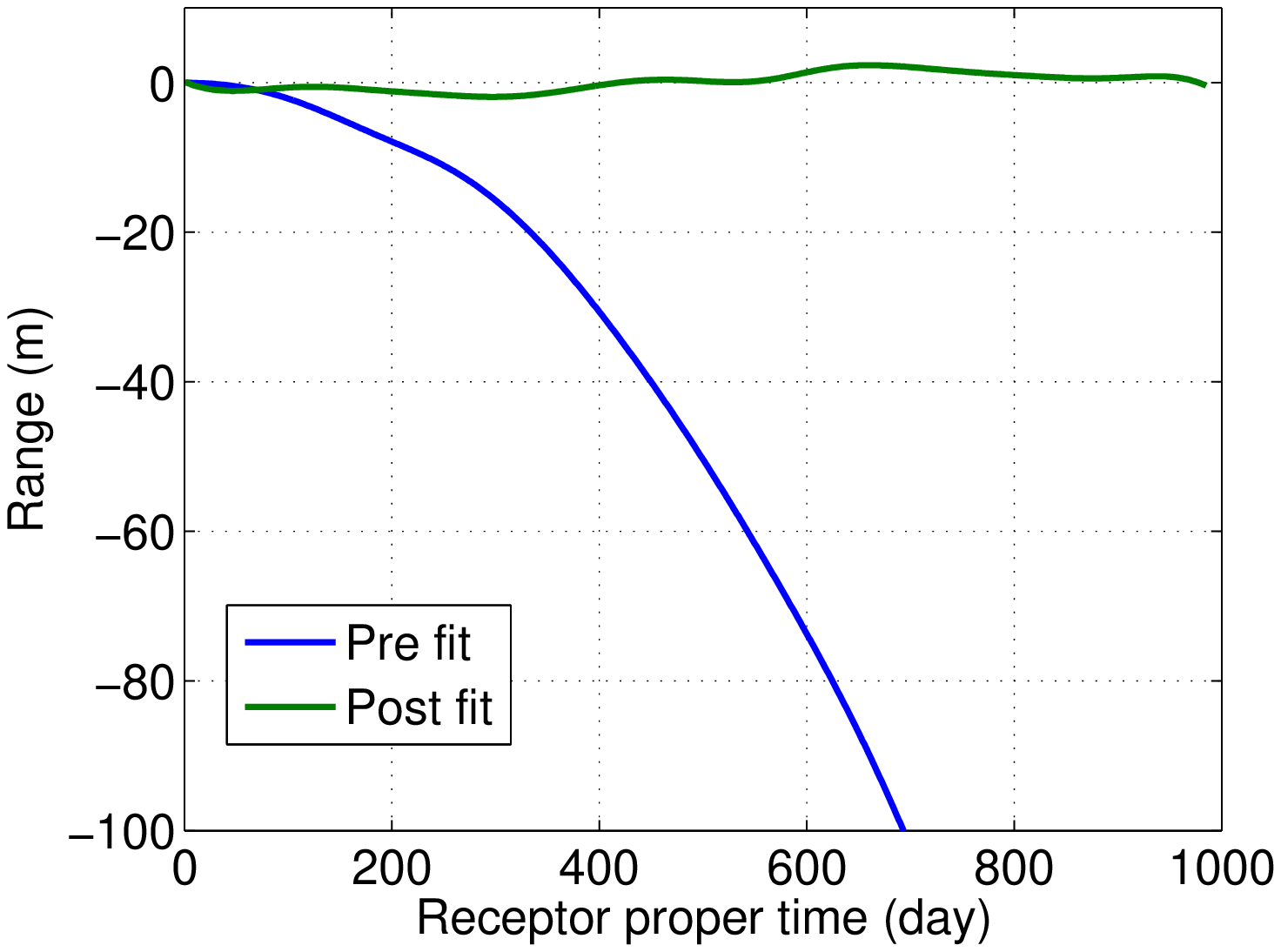,width=0.47\textwidth}\hfill
\psfig{figure=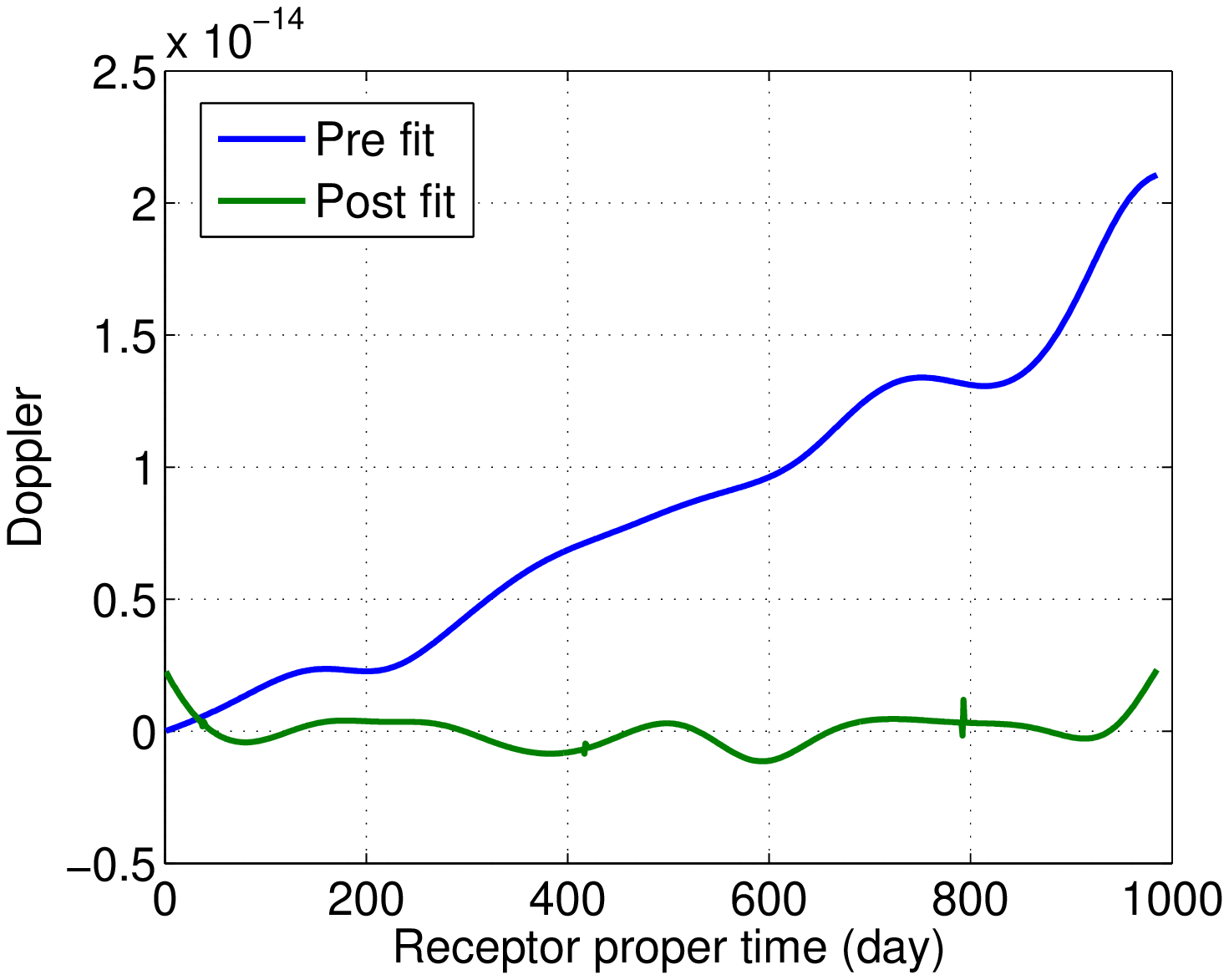,width=0.47\textwidth}
\caption{Representation of the Range (on the left) and Doppler (on the right) signals due to the MOND EFE ($Q_2=4.1 \ 10^{-26}s^{-2}$). The blue line is the difference between a simulation with the EFE and a simulation in GR (with the same parameters). The green line shows the residuals obtained after analyzing the simulated data in GR (which means after the fit of the different parameters).}
\label{figMond}
\end{figure}

Figure \ref{figMond} represents the effect of the EFE on the Range and Doppler signals from Cassini. It can be seen that the signal are just below the Cassini accuracy ($10^{-14}$ in Doppler). Therefore, the Cassini arc considered here is not sufficient to provide a good test of MOND External Field Effect.

\section{Conclusion}
In this communication, we have presented a new tool that performs Range/Doppler simulations in metric theories of gravity. With this software, it is easy to get the order of magnitude and the signature of the modifications induced by alternative theories of gravity on radioscience signals. As an example, we have presented some simulations for the Cassini mission in Post-Einsteinian Gravity and we have derived boundary values for some PEG parameters. We have also presented simulations including the MOND External Field Effect and we have shown that this effect is too small to be detected during the Cassini cruise between Jupiter and Saturn.

In the future, further simulations can be done for other theories and other (future and past) space missions. 

\section*{Acknowledgments}
A.~Hees is research fellow from FRS-FNRS (Belgian Fund for Scientific Research). Numerical simulations were made on the local computing ressources (Cluster URBM-SYSDYN) at the University of Namur (FUNDP, Belgium).

\end{document}